\documentclass[twocolumn]{IEEEtran} 

\usepackage{epsfig}
\def\BibTeX{{\rm B\kern-.05em{\sc i\kern-.025em b}\kern-.08em
    T\kern-.1667em\lower.7ex\hbox{E}\kern-.125emX}}
\newcommand{\Pb}{\mbox{P}}
\newcommand{\E}{\mbox{E}}

\begin{document}

\title{Uplink User Capacity in a CDMA System with Hotspot Microcells:  Effects
of Finite Transmit Power and Dispersion} 

\author{Shalinee Kishore, Larry J. Greenstein, H. Vincent Poor, Stuart C. Schwartz
\thanks{S. Kishore is with the Department of Electrical
and Computer Engineering, Lehigh University, Bethlehem, PA, USA. L.J.
Greenstein is with WINLAB, Rutgers University, Piscataway, NJ, USA. H.V. Poor
and S.C. Schwartz are with the Department of Electrical Engineering,
Princeton University, Princeton, NJ, USA.  This research was jointly
supported by the New Jersey Commission on Science and Technology, the
National Science Foundation under CAREER Grant CCF-03-46945, Grant
ANI-03-38807, and the AT\&T Labs Fellowship Program. Contact:
skishore@eecs.lehigh.edu.} }


\maketitle

\begin{abstract}
This paper examines the uplink user capacity in a two-tier code division
multiple access (CDMA) system with hotspot microcells when user terminal
power is limited and the wireless channel is {\em finitely-dispersive}.  A
finitely-dispersive channel causes variable fading of the signal power at the
output of the RAKE receiver.  First, a two-cell system composed of one
macrocell and one embedded microcell is studied and analytical methods are
developed to estimate the user capacity as a function of a dimensionless
parameter that depends on the transmit power constraint and cell radius.
Next, novel analytical methods are developed to study the effect of variable
fading, both with and without transmit power constraints.  Finally, the
analytical methods are extended to estimate uplink user capacity for {\em
multicell} CDMA systems, composed of multiple macrocells and multiple
embedded microcells.  In all cases, the analysis-based estimates are compared
with and confirmed by simulation results.
\end{abstract}

\section{Introduction}
Wireless operators often install low-power {\em hotspot microcell} base stations to
provide coverage to small high-traffic regions within a
larger coverage area.  These microcells enhance the
user capacity and coverage area supported by the existing
high-power {\em macrocell} base stations.  In this paper, we study these
gains for a {\em two-tier code division multiple access (CDMA) network} in which both the macrocells and microcells
use the same set of frequencies.  Specifically, we study the effects,
on the uplink capacity and coverage area, of transmit power
constraints and variable power fading due to multipath.
\\
\\
Earlier studies have examined the uplink performance of these
single-frequency, two-tier CDMA systems, e.g., \cite{Shapira}-\cite{skishore3}.
In \cite{skishore1}, we showed how to compute the uplink user capacity for a
single-macrocell/single-microcell (two-cell) system using exact and
approximate analytical methods that accounted for random user
locations, propagation effects, signal-to-interference-plus-noise ratio (SINR)-based power control, and various
methods of base station selection. In \cite{skishore2}, we calculated
capacity gains under similar assumptions for a system composed of
multiple macrocells and/or multiple microcells; the results pointed to a
roughly linear growth in capacity as the number of these base stations
increase, where the constants of the linear curve were derivable
solely from the analysis for the two-cell case.  This linear approximation was conjectured based on the trends observed from several simulations.  More recently, we have developed an {\em analytical} approximation to the user capacity of a multicell system which is much tighter than the empirical result of \cite{skishore2}; and is valid over a larger number of embedded microcells.  Interestingly, this new analytical approach also depends on constants obtained from a far-simpler two-cell analysis.  These studies therefore highlight the importance of understanding two-cell performance in computing the performance of larger multicell systems.
\\
\\
The capacity gains demonstrated in these earlier works
were based on
assumptions that 1) user terminals have unlimited power; and 2) that the
wireless channel is so wideband that user signals have
constant output power after RAKE receiver processing.  This latter condition is
equivalent to assuming that user signals go through an
infinitely-dispersive channel before reception.\footnote{By ``infinitely dispersive,'' we mean that the channel has an
infinitude of significant paths so that, after RAKE processing, the receiver signal output has constant power.}  In this paper, we remove these
conditions.  Portions of this work were presented in \cite{skishore4} and a companion paper on {\em downlink} capacity \cite{skishore5} showed that this type of system is uplink-limited.  Here, we improve the presentation in \cite{skishore4} and present several new results.  Specifically, we show how the capacity/coverage tradeoff under finite power constraints can vary significantly with shadow fading.  Further, we bridge our study of finite power constraints and finite channel dispersion by presenting a new analysis of user capacity in two-cell systems under {\em both} maximum transmit power constraints and variable power fading.  Finally and most significantly, we develop and verify analytical expressions for the total user capacity of {\em multicell} two-tier CDMA systems under finite dispersion.
\\
\\
Section \ref{sysdes} describes the basic two-cell system and the model used to capture the power gain between a user and a base; and it summarizes our previous results for total uplink capacity.  Section \ref{pmax_sec} assumes a limit on terminal transmit power and approximates total capacity as a function of this power and cell size.  Section \ref{varfading_sec} relaxes the condition on infinite dispersion and approximates the total capacity, for both limited and unlimited terminal power, as a function of the multipath delay profile.  Section \ref{multimulti_sec} extends the results for finite dispersion and unlimited terminal power to the case of multiple macrocells and multiple microcells.  Simulations confirm the accuracy of the approximation methods over a wide range of practical conditions and assumptions.
\section{System and Channel Description}
\label{sysdes}
{\bf The System:}  We first consider a system with a macrocell and an embedded microcell which together contain $N$ total users, comprising $N_M$ in the macrocell and
$N_\mu=N-N_\mu$ in the microcell.  We assume users have random codes of
length $W/R$, where $W$ is the system bandwidth and $R$ is the user rate.  Each user is power-controlled by its base so as to achieve a required uplink power level.  It was shown in \cite{skishore1} that the required received power
levels to meet a minimum SINR requirement of $\Gamma$ are
\begin{equation}
S_{M}= \eta W \frac{K-N_\mu+I_M}{(K-N_\mu)(K-N_M)-I_MI_\mu}
\label{pwr1}
\end{equation}
and
\begin{equation}
S_{\mu}= \eta W \frac{K-N_M+I_\mu}{(K-N_\mu)(K-N_M)-I_MI_\mu}
\label{pwr2}
\end{equation}
at the macrocell and microcell base stations, respectively. Here, $K =
1+(W/R)/\Gamma$ denotes the single-cell pole capacity
\cite{gilhousen}, $\eta$ is the power spectral density of the additive
white Gaussian noise, and $I_M$ and $I_\mu$ are {\em normalized cross-tier
interference} terms.  Specifically, $S_\mu I_M$ is the total interference power at the macrocell due to microcell users and $S_M I_\mu$ is the total interference power at the microcell due to macrocell users.  Thus,
\begin{equation}
I_M = \sum_{j \in U_\mu}\frac{T_{Mj}}{T_{\mu j}}; \mbox{~~~~~} I_\mu = \sum_{j \in U_M}\frac{T_{\mu j}}{T_{Mj}},
\label{IM_Imu}
\end{equation}
where $U_\mu$ represents the set of microcell users, $U_M$ represents the
set of macrocell users, and $T_{Mk}$ and $T_{\mu k}$ are the path
gains from user $k$ to the macrocell and microcell base stations,
respectively.  We say that a given arrangement of user locations and path gains is {\em feasible} if and only if the common denominator in (\ref{pwr1}) and (\ref{pwr2}) is positive. The fraction of all possible cases (i.e., all combinations of $I_M$ and $I_\mu$) for which this occurs is called the {\em probability of feasibility}.  The cross-tier interference terms depend on the path gains  for all users to both base stations and on the method by which users select base stations, leading to the sets $U_\mu$ and $U_M$.
\\
\\
{\bf Path Gain Model:} We model the instantaneous power gain (i.e., the sum of the power gains of the resolvable multipaths) between user $k$
and base station $l$ as
\begin{eqnarray}
T_{lk}=\left\{ \begin{array}{cc} H_l \left( \frac{b_l}{d_{lk}} \right)^2 10^{\zeta/10}\rho & d_{lk}
< b_l \\ H_l \left( \frac{b_l}{d_{lk}} \right)^4 10^{\zeta/10}\rho & d_{lk} \geq b_l \end{array}
\right.,
\end{eqnarray}
where $d_{lk}$ is the distance between user $k$ and base station $l$; $b_l$
is the breakpoint distance of base station $l$ at which the slope of the
decibel path gain versus distance changes; $H_l$ is a gain factor that captures the
effects of antenna height and gain at base station $l$;
$10^{\zeta/10}$ represents {\em shadow fading}; and $\rho$ is the variable
fading due to multipath \cite{Erceg}-\cite{rap}.  Note that $\zeta$ and $\rho$
are different for each user-base pair ($l,k$) but, for convenience, we suppress
the subscripts for these terms.  Since the antenna height and gain are greater
at the macrocell, we can assume $H_M > H_\mu$.  We model $\zeta$ as a zero
mean Gaussian random variable with variance $\sigma^2_l$ for base
station $l$. The fading due to multipath, $\rho$, is scaled to be a
unit-mean random variable ($\E \{ \rho \} =1$).  In an infinitely dispersive channel, $\rho
= 1$.  Let $\tilde{T}_{lk}$ denote the path gain
between user $k$ and base $l$ in an infinitely dispersive channel.
Thus, $\tilde{T}_{lk}$ represents the
local spatial average of
$T_{lk}$, and $T_{lk}=\tilde{T}_{lk} \rho$.
\\
\\
{\bf Base-Selection Scheme:} We assume a user $k$ chooses to
communicate with the macrocell base station if $\tilde{T}_{Mk}$ exceeds $\tilde{T}_{\mu k}$ by a fraction $\delta$, called the {\em desensitivity} parameter.  Thus, in both finitely-dispersive and infinitely
dispersive channels, the user
selects base stations according to the local mean path gains.\footnote{In
\cite{skishore1}, we studied this path gain-based selection method as
well as the selection scheme in which users are assigned to
base stations such that each terminal requires the least transmit power.
The result show that the latter scheme out-performs the path-gain-based method by roughly 15\% in capacity.}  With no loss in generality, we assume $\delta =1$ in this study.
\\
\\
{\bf Capacity:}  The path gain model and the selection scheme described above imply that the
cross-tier interference terms are random variables; they depend on the random variations due to shadow and multipath fading  and on the random user
locations.  In \cite{skishore1}, we demonstrated how to compute the
cumulative distribution functions (CDF's) of the two cross-tier
interference terms in (\ref{IM_Imu}) under the condition of infinite dispersion.  Let $\tilde{I}_M$
and $\tilde{I}_\mu$ denote these cross-tier terms under that condition.  The CDF's of $\tilde{I}_M$ and $\tilde{I}_\mu$ were used to compute the
exact uplink user capacity in a two-cell system with unlimited terminal
power.  In addition, the mean
values of $\tilde{I}_M$ and $\tilde{I}_\mu$ were used to reliably approximate attainable user capacity for the two-cell system as
\begin{equation}
\tilde{N}=\frac{2K}{1+\sqrt{\tilde{v}_M \tilde{v_\mu}}},
\label{Nunif}
\end{equation}
where $\tilde{v}_M$ and $\tilde{v}_\mu$ are the mean values (over all user
locations and shadowing fadings) of single
terms in the sums $\tilde{I}_M$ and $\tilde{I}_\mu$,
respectively.\footnote{In computing $\tilde{N}$, we used the fact, proved in \cite{skishore2}, that
maximal total capacity results when $N_M=N_\mu$.}  The total user
capacity of the two-cell system thus depends on $K$ (a system
parameter) and the product $\tilde{v}_M\tilde{v}_\mu$.  We
computed this product under various conditions and observed
that it (and therefore $\tilde{N}$) are robust to variations in propagation parameters, separation
between the two base stations, and user distribution.  Its value under
all conditions examined was about 0.09, leading to the robust result $\tilde{N} \approx 1.5K$.
\section{User Capacity with Limited Terminal Power}
\label{pmax_sec}
\subsection{Analysis}
\label{pmax_sec1}
In this section, we assume an infinitely dispersive channel and study
a system where each  user terminal has a maximum transmit power
constraint of $P_{\max}$.   The two-cell system is unable to support a
given set of $N$ users if the users are {\em infeasible} (i.e.,
unsupportable even with unlimited terminal power) or if any one of the
$N$ users requires a transmit power higher than $P_{\max}$.  We have
already computed the probability of infeasibility of $N$ total users
in \cite{skishore1}; we now determine the probability that any one of
$N$ feasible users exceeds the terminal power limit of $P_{\max}$.
The combined event, when the system is either infeasible or feasible
but a user terminal exceeds the transmit power requirements, is
referred to as {\em outage}.  The probability of outage for $N$ users
can then be written as
\begin{equation}
P_{out}(N) = P_{inf}(N) + (1-P_{inf}(N)) \cdot \mbox{Pr~}[P>P_{\max}|N],
\label{totaloutage}
\end{equation}
where $P_{inf}(N)$ is the probability of infeasibility of $N$ users
and $\mbox{Pr~}[P>P_{\max}|N]$ is the probability that the transmit power of
any one of the $N$ feasible users exceeds $P_{\max}$.  We seek the
values of $N$ that a two-cell system can support for a specified
probability of outage, i.e., we desire the largest $N$ such that $P_{out}(N) \leq
\alpha_{out}$.
\\
\\
We assume $b_M=b_\mu=b$.  For analytical purposes, it is convenient to
write the path gain between user $k$ and base station $l$ as:
\begin{equation}
\tilde{T}_{lk}=H_l \left( \frac{b}{d_{\max}} \right)^4 \tilde{T}'_{lk},
\end{equation}
where $d_{\max}$ is the desired maximum distance from the macrocell base station to a system user (it can be regarded as the macrocell radius), and
\begin{eqnarray}
\tilde{T}'_{lk}=\left\{ \begin{array}{cc} \left( \frac{d_{\max}}{d_{lk}} \right)^2 \left( \frac{d_{\max}}{b} \right)^2 10^{\zeta/10}, & d_{lk} < b \\
\left( \frac{d_{\max}}{d_{lk}} \right)^4 10^{\zeta/10}, & d_{lk} \geq b
\end{array} \right. \mbox{~.}
\label{mod_tgain}
\end{eqnarray}
Microcell user $j$ transmits at power $P=S_{\mu j}/\tilde{T}_{\mu j}$;
thus $P$ exceeds $P_{\max}$ if
\begin{equation}
\frac{\tilde{S}'_\mu}{\tilde{T}'_{\mu j}} > \frac{P_{\max} H_\mu}{\eta W}
\left( \frac{b}{d_{\max}} \right)^4 \label{peruser_outage_mu} \equiv F
\end{equation}
where $\tilde{S}'_\mu = S_\mu/\eta W$.  Similarly, the required transmit power of macrocell user $i$ exceeds $P_{\max}$ if
\begin{equation}
\frac{\tilde{S}_M'}{\tilde{T}_{Mi}} > \frac{P_{\max} H_M}{\eta W}
\left( \frac{b}{d_{\max}} \right)^4 = FH,
\end{equation}
where $\tilde{S}_M'=S_M/\eta W$ and $H=H_M/H_\mu$.
\\
\\
Note that $\tilde{S}_M$ and $\tilde{S}_\mu$ and all path gains are random variables, related to the randomness of user locations and shadow fadings.  Following
the steps outlined in \cite{skishore1} to obtain the probability distributions of
$\tilde{I}_M$ and $\tilde{I}_\mu$, we can extend the analysis to compute the
distributions of $\tilde{S}'_M$ and $\tilde{S}'_\mu$.  Next, we can determine the
distributions of $\tilde{T}'_M$ and $\tilde{T}'_\mu$ for a random
macrocell user and a random microcell user, respectively.  With the
distributions of $\tilde{S}'_M$, $\tilde{S}'_\mu$, $\tilde{T}'_{M}$ and
$\tilde{T}'_{\mu}$, we can then determine $\mbox{Pr~}[P>P_{\max}|N]$ as a function
of $F$:
\begin{eqnarray}
\mbox{Pr~} [P>P_{\max}|N]=\sum_{n=0}^N \left( \begin{array}{c} N \\ n \end{array} \right) p^nq^{N-n} (1-p_M^n p_\mu^{N-n}),
\label{outage_eq}
\end{eqnarray}
where $p$ is the probability of a user assignment to the
macrocell;\footnote{$p$ can be computed using the methods in
\cite{skishore1}.} $q=1-p$;
\begin{equation}
p_M = \Pb [\tilde{S}_M'/\tilde{T}_M' \leq FH]; \mbox{~~~~~}p_\mu = \Pb
[\tilde{S}_\mu'/\tilde{T}_\mu' \leq F].
\end{equation}
Finally, we can determine the largest $N$ for which $P_{out}(N) \leq \alpha_{out}$ for a given value of $F$.  However, instead of computing the distributions of $\tilde{S}'_M$ and
$\tilde{S}'_\mu$, we propose an approximate method that works quite
well, namely, using the means of $\tilde{S}'_M$ and $\tilde{S}'_\mu$.  This approach retains (\ref{totaloutage}) and (\ref{outage_eq}) but uses the following modified values for $p_M$ and
$p_\mu$:
\begin{eqnarray}
\tilde{p}'_M &=& \mbox{Pr~} [\mbox{E~} \{ \tilde{S}'_M \} /\tilde{T}'_{M}\leq FH]; \\
\tilde{p}'_\mu &=& \mbox{Pr~} [\mbox{E~} \{ \tilde{S}'_\mu \} /\tilde{T}'_{\mu} \leq F].
\end{eqnarray}
This greatly simplifies calculating the
probability of outage, with minor loss in accuracy.
\subsection{Numerical Results}
\begin{figure}[t]
\begin{center}
\epsfig{figure=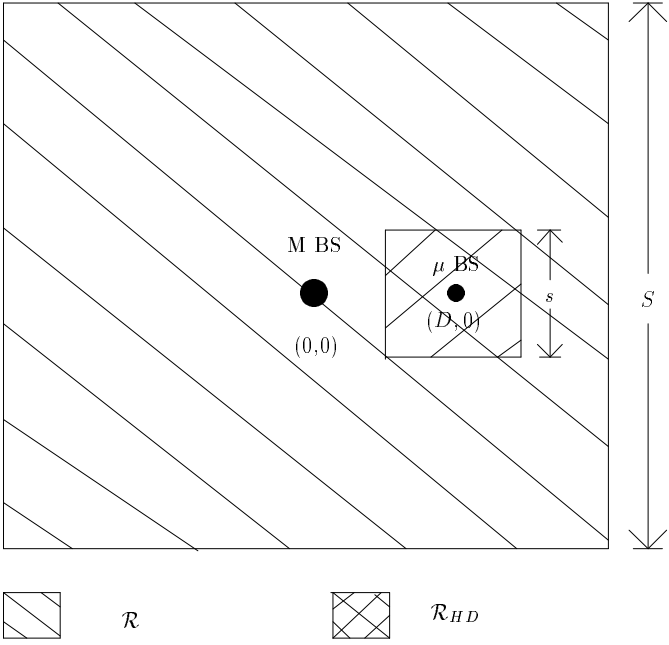,width=3.5in} \caption{A two-cell system composed of a larger square region with a smaller square region containing a high-density of users.} \label{fxy}
\end{center}
\end{figure}

\begin{figure}[t]
\begin{center}
\epsfig{figure=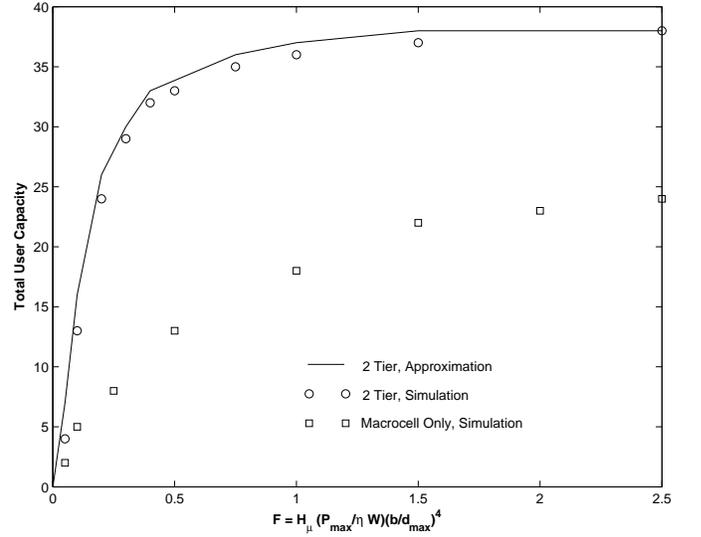,width=3.5in} \caption{Total user capacity as a function of $F$.  Results are for a single macrocell with or without an embedded microcell, with an analytical approximation shown for the former. (Infinitely dispersive channel)} \label{pwr_out1}
\end{center}
\end{figure}

\begin{figure}[t]
\begin{center}
\epsfig{figure=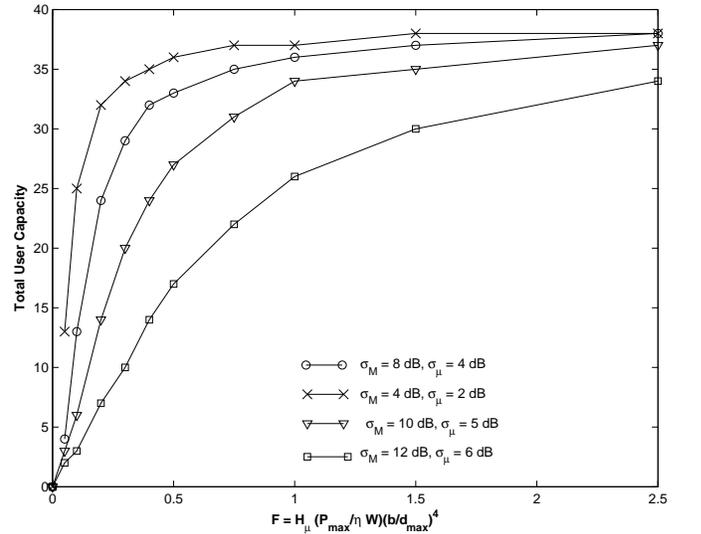,width=3.5in} \caption{Total user capacity as a function of $F$ for various combinations of
$\sigma_M$ and $\sigma_\mu$.  (Two-cell two-tier system and infinitely dispersive channel)} \label{sigma_curves}
\end{center}
\end{figure}

We use simulation to study the two-cell system in Figure \ref{fxy}.  Specifically, we assume a square geographic region, with sides of length $S$.  At a distance $D$ from the center of this square is a smaller square region, with sides of length $s$.  A portion of the total user population $N$ is uniformly distributed over the larger square region and the remaining users are uniformly distributed over the smaller square region.  The smaller square thus represents a traffic hotspot within the larger coverage region.  A macrocell base with antenna height $h_M$ is at the center of the larger square, while a microcell base with antenna height $h_\mu$ is at the center of the smaller square.  We assume each user terminal has antenna height $h_m$.  These antenna heights are used to compute the distance between transmit antenna at each user terminal and the receive antenna at the two base stations.  The breakpoint of the path loss to the microcell base is engineered to be $s/2$, to help ensure that the
microcells provide strong signals to those users contained within the
high density region, and sharply diminishing signals to users outside
it.  Finally, we assume that on average half of the $N$ users are uniformly distributed over the hotspot region surrounding the microcell base.  This is done to obtain a roughly equal
number of users for each base, i.e., to ensure that the
maximal user capacity occurs with high probability.
\\
\\
Using the parameter values listed in Table 1, we determine the total user capacity that can be sustained with 5\% outage probability for various values of $P_{\max}$.  Figure \ref{pwr_out1} shows the resulting user
capacity as a function of $F$ as defined in (\ref{peruser_outage_mu}).
We observe the close correspondence
between the approximate and simulation results.  We also note that the
user capacity goes to an asymptotic value as $F$ gets large and that,
below a critical value of $F$ (denoted by $F^*$), the user capacity
decreases sharply as $F$ decreases.  (From the figure, $F^* \approx 1$.)  In other words, there are critical combinations of $P_{\max}$
and $d_{\max}$ such that, if we either reduce the transmit power
constraint or increase the desired coverage area, the overall user
capacity noticeably drops; and, for values of $F>F^*$, the system
operates as if there is unlimited terminal power.
\\
\\
In an earlier study \cite{thesis}, we found that total user capacity is nearly invariant to $\sigma_M$ and $\sigma_\mu$, increasing {\em very} slightly as $\sigma_M$ and $\sigma_\mu$ increase.  This finding was for unlimited terminal power.  We now consider the
uplink user capacity as a function of $F$ for various pairs of
$(\sigma_M,\sigma_\mu)$.  The results, obtained via simulation, are in Figure \ref{sigma_curves}.
Even though the capacity for unlimited transmit power ($F$ infinite) is
slightly larger for $\sigma_M=12$,$\sigma_\mu=6$ \cite{thesis}, the curves increase much faster for smaller values of $\sigma_M$ and
$\sigma_\mu$.  Thus, although for unlimited
transmit power the total user capacity is roughly invariant to $\sigma_M$ and
$\sigma_\mu$, it can vary significantly
with these parameters under transmit power
constraints.
\section{User Capacity with Variable Fading:  Two-Cell System}
\label{varfading_sec}
\subsection{Analysis}
We now consider {\em finite dispersion} (i.e., a finite number of significant multipaths), which causes the sum of the multipath powers to fade with time as a user moves in the environment.  We assume that users still select bases according to the {\em slowly} varying path gains, $\tilde{T}_{lk}$, but the fluctuations of signals and interferences due to multipath lead to instantaneous occurrences of outage.
\subsubsection{Channel Delay Profile}
The instantaneous path gain from base $l$ to user $k$ is $T_{lk}=\rho \tilde{T}_{lk}$, where $\rho$ is independent and identically distributed (i.i.d.) over all $(l,k)$.  Let $r_n$ represent the instantaneous power gain of the $n$-th multipath component of a particular uplink channel.  Then, for that channel,
\begin{equation}
\rho=\sum_{n=1}^{L_p} r_n \label{rho_def},
\end{equation}
where $L_p$ is the total number of multipaths and we assume a scaling such that $\E \{ \rho \} =1$.  We consider four possible multipath delay profiles, for each of which $\rho$ has a different statistical nature.  One is for the so-called {\em uniform channel}, in which the $L_p$ multipaths are i.i.d. and Rayleigh-fading, i.e., each has a power gain that is exponentially distributed with a mean of $1/L_p$.
\\
\\
The other three delay profiles studied here are based on
cellular channel models for the typical urban (TU), rural area (RA),
and hilly terrain (HT) environments provided in third-generation standards. See, for example,
\cite{3gpp}, which tabulates $\E\{ r_n \}$ in dB for these environments.

\subsubsection{Methodology}
Base selections (and thus, the sets $U_\mu$ and $U_M$) are made according
to the local average path gains, $\tilde{T}_{lk}$.  The instantaneous
cross-tier interference powers are
\begin{eqnarray}
I_M&=&\sum_{i \in U_\mu} \frac{T_{Mi}}{T_{\mu i}} = \sum_{i \in U_\mu} \frac{\rho_{Mi}}{\rho_{\mu i}} \cdot \frac{\tilde{T}_{Mi}}{\tilde{T}_{\mu i}}; \label{IM_new}\\
I_\mu&=&\sum_{i \in U_M} \frac{T_{\mu i}}{T_{Mi}} = \sum_{i \in U_M} \frac{\rho_{\mu i}}{\rho_{M i}} \cdot \frac{\tilde{T}_{\mu i}}{\tilde{T}_{Mi}} \label{Imu_new}\end{eqnarray}
where $\rho_{Mi}$ and $\rho_{\mu i}$ are i.i.d. random variables,
each distributed as $\rho$ in (\ref{rho_def}).  We see from the definitions
of $I_M$ and $I_\mu$ that, at any one instant, the
system could become infeasible or, more generally, experience outage,
depending on the values of $\rho_{Mi}$ and $\rho_{\mu i}$.  Here again, we seek the largest number of users, $N$, that can be supported such that $P_{out}(N) \leq  \alpha_{out}$.
\\
\\
The exact method for calculating $N$ requires that the distributions of $I_M$ and $I_\mu$ be computed.  The procedure is even more complex than before since we must incorporate the effects of the additional random quantities, $\rho_{Mi}$ and $\rho_{\mu i}$.  Thus, we resort to two analytical approximations.  The first estimates total user capacity under finite dispersion but for large $F$ (i.e., $F > F^*$); and the second estimates capacity under both finite dispersion and finite transmit power requirements.

\subsubsection{Total User Capacity for $F$ Large}
\label{fading_f_large}

\paragraph{Estimated User Capacity in a Uniform Channel}  The total user capacity in a uniform channel, $N_u$, can be approximated by modifying the mean method presented in (\ref{Nunif}).  Specifically, we postulate that
\begin{equation}
N_u = \frac{2K}{1+\sqrt{\overline{v}_M \overline{v}_\mu}},
\label{Nfading_approx}
\end{equation}
where $\overline{v}_M=\E\{ \kappa \} \tilde{v}_M$ and $\overline{v}_\mu=\E \{ \kappa \} \tilde{v}_\mu.$  The mean values $\tilde{v}_M$
and $\tilde{v}_\mu$ were defined earlier; and $\kappa=\rho_1/\rho_2$, where $\rho_1$ and $\rho_2$ are i.i.d. Gamma
random variables with unit mean and $L_p$ degrees of freedom.  In other words, $\rho_1$
represents either $\rho_{Mi}$ or $\rho_{\mu i}$, and $\rho_2$ represents
the other. For a uniform channel with $L_p$ paths, the probability density of $\rho$ (and therefore of $\rho_1$ or $\rho_2$) is
\begin{equation}
f_\rho (x) = \frac{L_p}{(L_p-1)!}(x L_p)^{L_p-1}e^{-x L_p}, \mbox{~~~}x>0.
\label{rho_density}
\end{equation}
We can compute the mean of $\kappa$ as $\E \{ \kappa \} = \E \{ \rho_1 \} \cdot \E \left\{ \rho_2^{-1} \right\}$, where $\E \{\rho_1 \}=1$ and
\begin{eqnarray}
\E \left\{ \frac{1}{\rho_2} \right\} &=&  \frac{L_p}{L_p-1} \int_0^{\infty} \frac{1}{(L_p-2)!} y^{L_p-2}e^{-y} \,dy \\
&=& \frac{L_p}{L_p-1} \mbox{,~~~~~}L_p>1.
\end{eqnarray}
We can therefore relate $N_u$ to $L_p$, as follows:
\begin{equation}
N_u(L_p) = \frac{2K}{1+\frac{L_p}{L_p-1}\sqrt{\tilde{v}_M \tilde{v}_\mu}}.
\label{Nu_approx}
\end{equation}
Clearly, this approximation breaks down as $L_p$ approaches 1, as it
predicts zero capacity in that case.\footnote{The true capacity for a
single-path Rayleigh fading channel is in fact quite poor, though
not zero.}  For $L_p>1$, we obtain a simple relationship between user
capacity and the degree of multipath dispersion.  As $L_p$ becomes infinite, user capacity converges to the estimate given by (\ref{Nunif}).
\paragraph{Estimated User Capacity in a Non-Uniform Channel}  The result in (\ref{Nu_approx}) can be used to approximate user capacity for {\em any} delay profile.  Consider a channel delay profile having some arbitrary variation of $\E \{ r_n \}$ with $n$ over the $L_p$ paths.  We can compute a {\em diversity factor}, $DF$, defined as the ratio of the square of the mean to the variance of $\rho$ in (\ref{rho_def}).  For independent Rayleigh paths, $DF$ can be shown to be \cite{lola}
\begin{equation}
DF = \frac{\left( \sum_{n=1}^{L_p} \E \{r_n\} \right)^2}{\sum_{n=1}^{L_p} (\E \{r_n \})^2}.
\label{DF_definition}
\end{equation}
From this definition, we see that the diversity factor for a uniform
channel is precisely $L_p$.\footnote{Furthermore, via a simple implementation of the Schwarz inequality, it can be shown that the diversity factor for a non-uniform channel with $L_p$ paths is less than $L_p$.}  The proposed approximation makes use of
this fact by first calculating the diversity factor for a given
(non-uniform) channel.  The approximate user capacity is then the value of $N_u$ corresponding to $L_p=DF$, i.e., $N_u(DF)$ in (\ref{Nu_approx}).
\subsubsection{Total User Capacity with Finite Terminal Power}
\label{fading_f_small}
To approximate user capacity under both finite dispersion and finite terminal power, we incorporate the approximations of Sections \ref{pmax_sec} and \ref{fading_f_large}.  We begin by first studying the uniform delay profile and then develop an approximation for a general non-uniform environment.
\paragraph{Estimated User Capacity in a Uniform Channel}  As in Section \ref{pmax_sec}, we seek the largest $N$ such that the probability of outage is $\alpha_{out}$ or lower.  In calculating outage, we use the expressions in (\ref{totaloutage}) and (\ref{outage_eq}) but replace $p_M$ and $p_\mu$ with $p'_M$ and $p'_\mu$, respectively, so that
\begin{eqnarray}
p'_M &=& \mbox{Pr~}[\mbox{E~}\{ S'_M \} /\rho \tilde{T}'_M \leq FH];\\
p'_\mu &=& \mbox{Pr~}[\mbox{E~} \{ S'_\mu \} / \rho \tilde{T}'_\mu \leq F], \label{pMmu_new}
\end{eqnarray}
where $\mbox{E~} \{ S'_M \} = \mbox{E~} \{ S_M \} / \eta W$, $\mbox{E~} \{ S'_\mu \} = \mbox{E~} \{S_\mu \} / \eta W$, and $\rho$ is a random variable with density given in (\ref{rho_density}).  The steps used to compute $\E \{S'_M \}$ and $\E \{ S'_\mu \}$ are identical to the steps outlined in Section \ref{pmax_sec} for the calculations of $\E \{ \tilde{S}'_M \}$ and $\E \{ \tilde{S}'_\mu \}$.  The difference here is that the densities of $S'_M$ and $S'_\mu$ depend on the densities of $I_M$ and $I_\mu$, (\ref{IM_Imu}), which account for variable fading, whereas $\tilde{S}'_M$ and $\tilde{S}'_\mu$ were obtained using the cross-tier interference terms in infinitely-dispersive channels.  The densities of $I_M$ and $I_\mu$ depend on the densities of the individual terms in their sums.  The terms in each sum are i.i.d. random variates, denoted here by $v_M$ and $v_\mu$.
\\
\\
For a uniform channel with $L_p$ paths, we note that $v_M = \kappa \tilde{v}_M$, where $\kappa = \rho_1 / \rho_2$, and $\rho_l$ ($l \in {1,2}$) has probability density given in (\ref{rho_density}).  Thus, the density of $v_j$ ($j \in \{ M, \mu \}$) can be computed as
\begin{equation}
f_{v_j}(v) = \int_{0}^{\infty} f_{\tilde{v}_j}(v/x)\cdot f_{\kappa}(x) \,dx, \mbox{~~~}v_j \in \{ M, \mu \},
\end{equation}
where the probability density of $\tilde{v}_j$ is given in \cite{skishore1} and
\begin{equation}
f_{\kappa}(x) = \frac{1}{(L_p-1)!} \sum_{i=0}^{L_p-1} \frac{(L_p-1+i)!}{i!} \cdot \frac{x^{i-1}(L_px-i)}{(x+1)^{L_p+1+i}}.
\end{equation}
We obtain $f_{\kappa}(x)$ from its CDF, which can be computed as\footnote{We use the following to compute the CDF of $\kappa$:  $\int_0^c \frac{x^m}{m!}e^{-x}\,dx = 1 - e^{-c} \sum_{i=0}^m \frac{c^i}{i!}$ and $\int_0^\infty x^n e^{- \alpha x} \,dx = n!/\alpha^{n+1}$.}\begin{eqnarray}
F_{\kappa}(x) &=& \Pb [\kappa \leq x] = \Pb [\rho_1 \leq \rho_2 x] \\
&=& 1-\frac{1}{(L_p-1)!} \sum_{i=0}^{L_p-1} \frac{x^i}{i!} \cdot \frac{(L_p-1+i)!}{(x+1)^{L_p+i}}.
\end{eqnarray}
With the density of $v_M$ and $v_\mu$ at hand, we use the analysis in \cite{skishore1} to compute the densities of $I_M$ and $I_\mu$ for a given value of $N$.  We then use the densities of these cross-tier interferences to compute the densities of $S'_M$ and $S'_\mu$, which in turn give us the mean values needed in (\ref{pMmu_new}).   The outage probability for this capacity $N$ is then determined using (\ref{totaloutage}) and (\ref{outage_eq}) with the substitutions for $p_M$ and $p_\mu$ indicated in (\ref{pMmu_new}).  Finally, we can approximate the maximum number of users supported for a desired outage level ($\alpha_{out}$) for the uniform multipath channel.
\paragraph{Estimated User Capacity in a Non-Uniform Channel}  Given a non-uniform channel, we propose an approximation to user capacity for a given value of $F$
that uses the above technique for the uniform channel.  Specifically, we use the results of Section \ref{fading_f_large} to claim that a system in a non-uniform channel with diversity factor $DF$ supports roughly the same number of users as the same system in a uniform channel with $DF$ paths.  Thus, for a given non-uniform delay profile, we first compute $DF$ from (\ref{DF_definition}).  We then set $L_p$ as the integer closest to $DF$ and use the analysis above to estimate the total user capacity with outage $\alpha_{out}$ in a uniform channel with $L_p$ paths.  This value approximates user capacity in the given non-uniform channel.

\subsection{Numerical Results}
\begin{figure}[t]
\begin{center}
\epsfig{figure=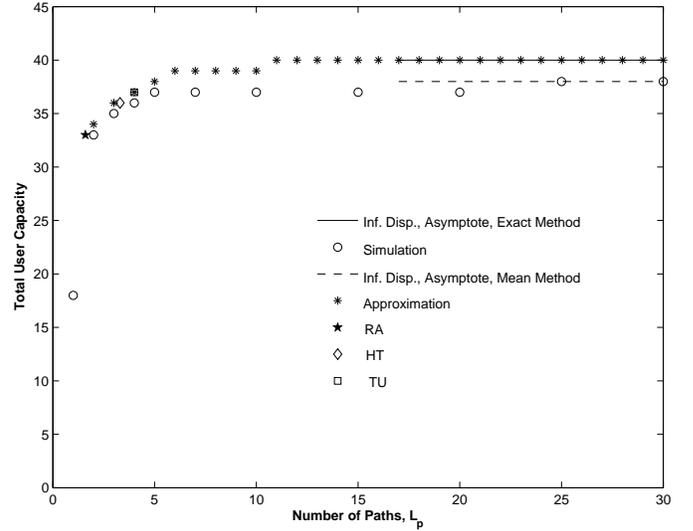,width=3.5in}
\caption{Total user capacity as a function of
$L_p$, using both simulation and analytical approximation.  Simulation results are also given for RA, HT, and TU channels, each at a value of $L_p$ equal to the channel's diversity factor.  (Two-cell two-tier system with unlimited terminal power)} \label{NvsL_fading}
\end{center}
\end{figure}

\begin{figure}[t]
\begin{center}
\epsfig{figure=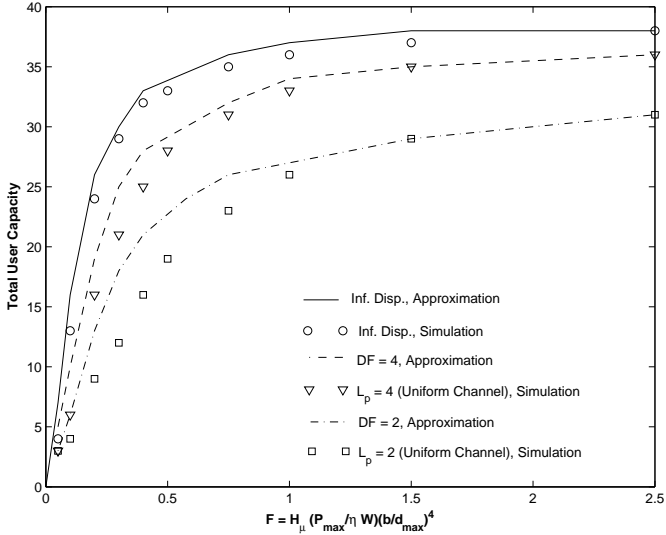,width=3.5in} \caption{Total user capacity for two uniform channels and the infinitely dispersive channel as a function of $F$, using both simulation and analytical approximation.  (Two-cell two-tier system)} \label{NvsF_fading}
\end{center}
\end{figure}

\begin{figure}[t]
\begin{center}
\epsfig{figure=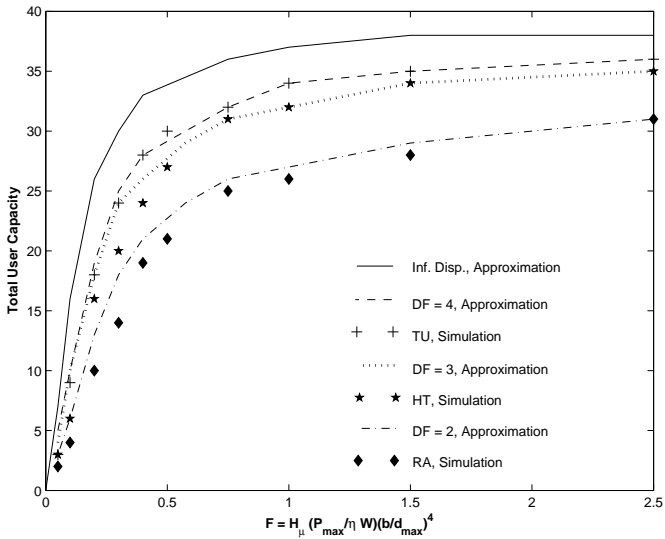,width=3.5in} \caption{Total user
capacity for RA, HT, TU, and infinitely dispersive channels as a function of $F$, using both simulation and analytical approximation.  (Two-cell two-tier system)} \label{NvsF_fading_2}
\end{center}
\end{figure}

We use the two-cell system described in Figure \ref{fxy} and Table 1 to obtain simulation results for finitely-dispersive channels.  For a given configuration of $N$ users, i.e., a fixed set of user locations and shadow fadings, we generate 1000 values of $\rho_M$ and $\rho_\mu$ for each user (based on the given channel delay profile) and record the instances of outage.  We then calculate this outage measurement over 999 other random user locations and shadow fadings (assuming a fixed value of $N$).  The total instances of outage over all these trials determine the probability of outage for $N$.
\\
\\
Using this simulation method, we first calculate the user capacity with 5\% outage probability when there are no transmit power constraints ($F=\infty$).\footnote{For $F$
large, 5\% outage is the same as 5\% infeasibility.}  This is done
for the uniform channel and the RA, HT, and TU environments.  For the uniform channel, we determine user capacity as a function of $L_p$;  Figure
\ref{NvsL_fading} shows results for both the
simulation and approximation methods.  The approximation curve follows
the simulation curve closely, especially for smaller values of
$L_p$.  Figure
\ref{NvsL_fading} also shows the infinite-$F$ capacity for the RA,
HT, and TU environments (from simulation).  We show each of these
capacities at a value of $L_p$ equal to its diversity factor.
The diversity factors for the RA, HT, and TU environments are 1.6,
3.3, and 4.0, respectively.  We see that, using the diversity factor, the capacity
of a non-uniform channel can be mapped to the computed curve for the
uniform channel, yielding a simple and reliable analytical
approximation to user capacity.  For channels with $DF > 5$, the
total user capacity appears to be within 5\% of that for an infinitely dispersive channel.
\\
\\
We also obtained user capacity (with 5\% outage probability) as a function of $F$ for the 2-path and 4-path uniform channels. This was done via both simulation and analytical approximation, and the
results are presented in Figure \ref{NvsF_fading}.  We observe, first, that the approximation matches the simulation results for the 2-path and 4-path channels in the region of interest, i.e., for $N \geq K$.  Overall, the approximations improve as $L_p$ increases.  Furthermore, the user capacities noticeably decrease for $F < F^*$, where $F^* \approx 1$ for all three channels studied here, a result which is consistent with the infinitely-dispersive case (Figure \ref{pwr_out1}).
\\
\\
In Figure \ref{NvsF_fading_2}, we plot user capacity versus $F$ using the approximation method of Section \ref{fading_f_small} for $L_p = 2,3,$ and 4 and simulation results for the RA, HT, and TU environments.  The RA results closely match those for $L_p=2$; the HT results closely match those for $L_p=3$; and the TU results closely match those for $L_p=4$.  Thus, the diversity factor method of Section \ref{fading_f_small} allows us to reliably estimate user capacity versus $F$ for any realistic delay profile.
\\
\\
Finally, we note that as in Figure \ref{NvsF_fading}, the critical value $F^*$ is around 1 for all cases in Figure \ref{NvsF_fading_2}.  Thus, we find that, regardless of the delay profile, $F^* \approx 1$.  For the remainder of the paper, we assume practical values of $P_{\max}$ and $d_{\max}$ such that $F > 1$.

\section{User Capacity with Variable Fading:  Multicell Systems}
\label{multimulti_sec}
\subsection{Estimated User Capacity in Infinitely-Dispersive Channels}
\label{multi_multi_inf_subsec}
In computing the approximate user capacity in an infinitely dispersive channel, we use a result in \cite{skishore2} which states that maximal user capacity results when each cell in a multicell system supports an equal number of users.  Let us denote $N_p$ as the (equal) number of users supported in each cell and $N_{\infty}(L,M)$ as the total number of users
supported in a system with $L$ microcells, $M$ macrocells, and an infinitely dispersive channel.\footnote{Note that $N_{\infty}(L,M)=(L+M)N_p$.}  We
assume initially that all $L$ microcells are embedded within one of the
$M$ macrocells (macrocell 1) and that the remaining $M-1$ macrocells surround this macrocell.  We first write $L+M$ inequalities
representing the minimum SINR requirements at the $L+M$ bases.  The received SINR's are determined using average interference terms.  For
example, the SINR requirement at macrocell base station 1 (which contains embedded microcells) is
\begin{equation}
\frac{\frac{W}{R} S_{M 1}}{N'_p S_{M1} + \sum_{l=1}^L S_{\mu l} N_p
\tilde{v}_M+\sum_{k =2}^M S_{M k} N_p \tilde{v}_{MM} + \eta W} \geq \Gamma ,
\label{multi_sinr_1}
\end{equation}
where $N'_P=N_p-1$, $S_{M j}$ is the required received power at macrocell $j$,
$N_p=N(L,M)/(L+M)$ (as per the optimal equal distribution of
users), and the same-tier interference between macrocells is represented by its average value, i.e., $N_p \tilde{v}_{MM}$, which we assume to be the same for all macrocells.  Here $\tilde{v}_{MM}$ is the interference into a macrocell caused by a user communicating with a neighboring macrocell, averaged over all possible user locations in $\mathcal R$ and over shadow fading.  In our calculations, we simplify the SINR requirement in (\ref{multi_sinr_1}) by assuming that $\tilde{v}_{MM} \approx \tilde{v}_M$, i.e., interference due to two users in neighboring macrocells is roughly equal to the interference at a macrocell due to an embedded microcell user.  Although macrocells transmit at higher powers than microcells, they are further apart from each other than is a macrocell base from an embedded microcell user.  This approximation assumes that the larger distance between macrocells balances the impact of higher transmit powers.
\\
\\
Next, we determine the SINR requirement at macrocell $j$ ($j \geq 2$).  To do so, we assume (as above) that $\tilde{v}_{MM} \approx \tilde{v}_M$.  We further assume that the interference from neighboring macrocells is much larger than the interference caused by transmissions in a few isolated microcells in a nearby macrocell, i.e., $\sum_{k \neq k}S_{Mk} \tilde{v}_{MM} N_p \gg \sum_{l=1}^L S_{\mu l}\tilde{v}_{M \mu} N_p$, where $\tilde{v}_{M \mu}$ is the interference at any macrocell base due to a microcell user embedded in a neighboring macrocell averaged over all possible user locations and shadow fadings. Under these two assumptions, the SINR requirement at macrocell $j$ ($j \geq 2$) is approximately
\begin{equation}
\frac{\frac{W}{R}S_{Mj}}{N'_pS_{Mj}+\sum_{k \neq j}S_{Mk}N_P \tilde{v}_{M} +\eta W} \geq \Gamma.
\end{equation}
Finally, we calculate the SINR requirements at the $L$ microcell bases.  Here, we assume first that $\tilde{v}_{\mu M} \approx \tilde{v}_{\mu}$, where $\tilde{v}_{\mu M}$ represents the interference into any microcell base due to a user in a neighboring macrocell averaged over user location and shadow fading.  In other words, we assume the interference at a microcell from neighboring macrocells is comparable to interference caused by transmission to the umbrella macrocell.  Note that this is a somewhat pessimistic assumption as $\tilde{v}_{\mu M}$ will typically be less than $\tilde{v}_\mu$.  This pessimistic approach is balanced by assuming further that the microcell-to-microcell interference $\tilde{v}_{\mu \mu}$ is negligible.  Here $\tilde{v}_{\mu \mu}$ is the average interference into any microcell base due to a user communicating with any other microcell.  Thus, the required SINR equation at microcell $l$ is
\begin{equation}
\frac{\frac{W}{R}S_{\mu l}}{N'_pS_{\mu l} + S_{M1} N_p \tilde{v}_\mu +
\sum_{k =2}^M S_{Mk}N_p \tilde{v}_{\mu} + \eta W} \geq
\Gamma.
\end{equation}
These $L+M$ inequalities can be used to show \cite{skishore3} that $S_{Mj} \geq 0$ and $S_{\mu l} \geq 0$ for all $j \in \{ 1,2,\ldots M \}$ and $l \in \{1,2, \ldots L \}$ if and only if
\begin{equation}
N_{\infty}(L,M) = (L+M)N_p \leq \frac{K (L+M)}{1+\sqrt{(L+M-1) \tilde{v}_M \tilde{v}_\mu}}.
\label{eq:cap_multi}
\end{equation}
A benefit of this simplified capacity condition is that it allows for capacity calculation using
$\tilde{v}_M$ and $\tilde{v}_\mu$ alone.  As a result, it is no more complex to
determine than is $N_T$, the user capacity when $M=1$ and $L=1$, Section 2.  Further, as the product $\tilde{v}_M \tilde{v}_\mu$ is robust to variations in propagation parameters and user distributions, so is the approximation in (\ref{eq:cap_multi}).  Despite its simplified form, this approximation yields a fairly
accurate estimate to the attainable user capacity.
\\
\\
A simple extension of the capacity expression in (\ref{eq:cap_multi}) can now be used to find the user capacity in a
{\em general} multicell system, i.e., a system in which the $L$ microcells can be embedded
anywhere within the coverage areas of the $M$ macrocells.  We first note that, in
the general system, each macrocell contains {\em on average} $L/M$
microcells.  Next, we use (\ref{eq:cap_multi}) to determine the user capacity for a multicell
system in which there are $M$ macrocells but only the center macrocell
contains $L/M$ microcells.  In
this case, each cell contains, $N_p$ users, where
\begin{equation}
N_p=\frac{K}{1+\sqrt{(\frac{L}{M}+M-1) \tilde{v}_M
    \tilde{v}_\mu}}.
\label{eq:cap_multi_2}
\end{equation}
Next, we assume that this value of $N_p$ hardly changes when some other macrocell contains (on average) $L/M$ embedded microcells.  Thus, we can finally approximate the total attainable capacity for the
general multicell system as $(L+M)$ times the result in (\ref{eq:cap_multi_2}), i.e.,
\begin{equation}
N_{\infty}(L,M) \approx \frac{K (L+M)}{1+\sqrt{(\frac{L}{M}+M-1) \tilde{v}_M
    \tilde{v}_\mu}}.
\label{eq:cap_multi_3}
\end{equation}
\subsection{Estimated User Capacity with Variable Fading}
The goal here is to develop a calculation for the multicell user capacity in {\em finitely dispersive channels} which is as straightforward to calculate as the capacity expression in (\ref{eq:cap_multi_3}).  Although there are several approaches to this problem, we propose an extremely simple and highly accurate approximation method.  First, we compute the diversity factor for the given multipath channel.  Next, we make the assumption that systems with the same diversity factor support equivalent numbers of users and we seek to find $N_{DF}(L,M)$, the user capacity in a system with $L$ microcells, $M$ macrocells, and a multipath channel with diversity factor $DF$.  We then use the results of Section \ref{varfading_sec} to compute the total user capacity for a two-cell system, i.e,
\begin{equation}
N_{DF}(1,1) = \frac{2K}{1+\frac{DF}{DF-1}\sqrt{\tilde{v}_M \tilde{v}_\mu}}.
\end{equation}
We also compute the total user capacity for a two-cell system with infinite dispersion, i.e.,
\begin{equation}
N_{\infty}(1,1) = \frac{2K}{1+\sqrt{\tilde{v}_M \tilde{v}_\mu}}.
\end{equation}
We use these two capacities to determine the capacity percentage loss due to dispersion, $p_{loss}(DF)$, for the two-cell system.  This loss is given as
\begin{equation}
p_{loss} = \frac{N_{DF}(1,1)}{N_{\infty}(1,1)} = \frac{1+\sqrt{\tilde{v}_M \tilde{v}_\mu }}{1 + \frac{DF}{DF-1} \sqrt{\tilde{v}_M \tilde{v}_\mu }}.
\end{equation}
Finally, we claim and demonstrate through extensive numerical results that the total attainable user capacity in a multicell system with $L$ microcells, $M$ macrocells, and finite dispersion $DF$ is simply $p_{loss}(DF)$ times the total attainable user capacity of a system with $L$ microcells, $M$ macrocells, and infinite dispersion, i.e.,
\begin{eqnarray}
N_{DF}(L,M) \approx p_{loss}(DF) \cdot N_{\infty}(L,M) \qquad \qquad \\
= \frac{1+\sqrt{\tilde{v}_M \tilde{v}_\mu }}{1 + \frac{DF}{DF-1} \sqrt{\tilde{v}_M \tilde{v}_\mu }} \cdot \frac{K(L+M)}{1+\sqrt{\left( \frac{L}{M} + M - 1 \right)\tilde{v}_M \tilde{v}_\mu }}.
\label{multi_fading_capacity}
\end{eqnarray}

\subsection{Numerical Results}
\subsubsection{Single-Macrocell/Multiple-Microcell System}
\begin{figure}[t]
\begin{center}
\epsfig{figure=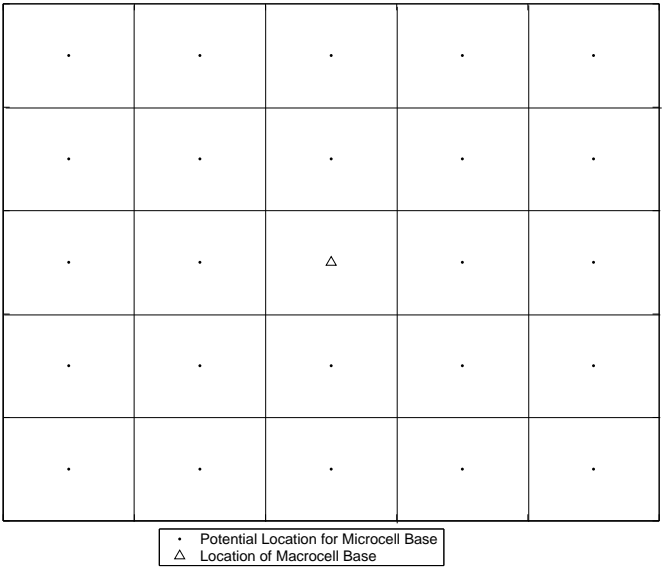,width=3.5in}
\caption{An example of the
single-macrocell/multiple-microcell system, with $n=5$.} \label{singlesquares}
\end{center}
\end{figure}

\begin{figure}[t]
\begin{center}
\epsfig{figure=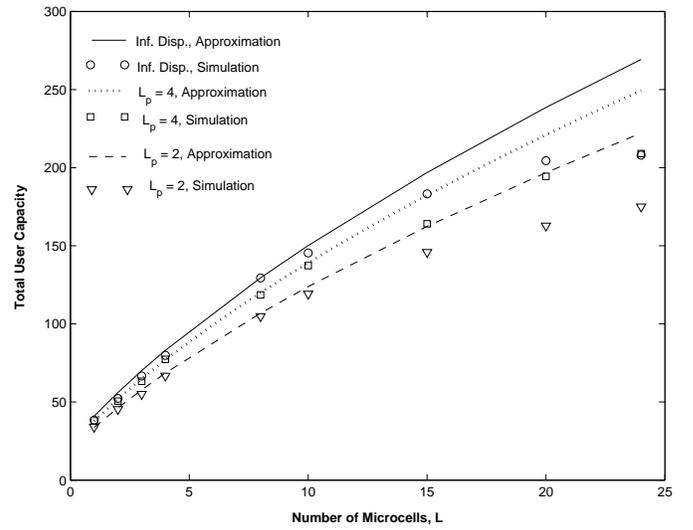,width=3.5in} \caption{Total user capacity as a function of $L$ for the 2-path and 4-path uniform channels and the infinitely dispersive channel using both simulation and analytical approximation.  Although not shown here, the density of user capacity obtained from simulation is fairly tight about the reported mean value.  (Unlimited terminal power)}
\label{ul_multicell_1a}
\end{center}
\end{figure}

We begin with a large square region having side $S$ and a
macrocell base at its center.   A fraction of the total
system users are uniformly distributed over this region.  This larger
square is divided into $n^2$ squares with side $s$, as shown in Figure
\ref{singlesquares} for $n=5$.  The smaller
squares represent potential high-density
regions. In each simulation trial, we randomly select $L$ high density
regions (excluding the center
square which contains the macrocell base) and place a microcell base at the center.  Finally, we assume that the $L$ microcells have identical
values for $b_\mu$, $H_\mu$, and $\sigma_\mu$.
\\
\\
Our simulations assume $n=5$, meaning 24 possible candidate high density regions and the parameters in Table 1.  For a given value of $L$, the simulation randomly selects $L$ of
the 24 high density regions.  A portion of the total user population
is uniformly-distributed over the large square region and
the remaining users are uniformly distributed over the selected
high density regions.  The average fraction of the total user
population placed in each region is $\frac{1}{L+1}$.  As before, this was done to ensure that maximal user capacity occurs with high probability \cite{skishore2}-\cite{skishore3}.  The simulation
then determines the total number
of users supported with 5\% outage.  This is done for 23 other random selections of $L$ high density regions, and the average over these selections is computed as a function of
$L$.
\\
\\
We performed the above simulations for uniform
channels with $L_p=2$ and $L_p = 4$, and for the infinitely dispersive channel.  The results are in Figure
\ref{ul_multicell_1a}.
In addition, we show the attainable user capacities predicted by
(\ref{eq:cap_multi_3}) and (\ref{multi_fading_capacity}).  The simulation points match the corresponding approximations very tightly up to about $L=12$, corresponding to a microcell {\em fill factor} (the fraction of macrocellular area served by microcells) of about one half.  In other words, our approximation technique works reliably in the domain {\em hotspot} microcells.
\subsubsection{Multiple-Macrocell/Multiple-Microcell System}
\begin{figure}[t]
\begin{center}
\epsfig{figure=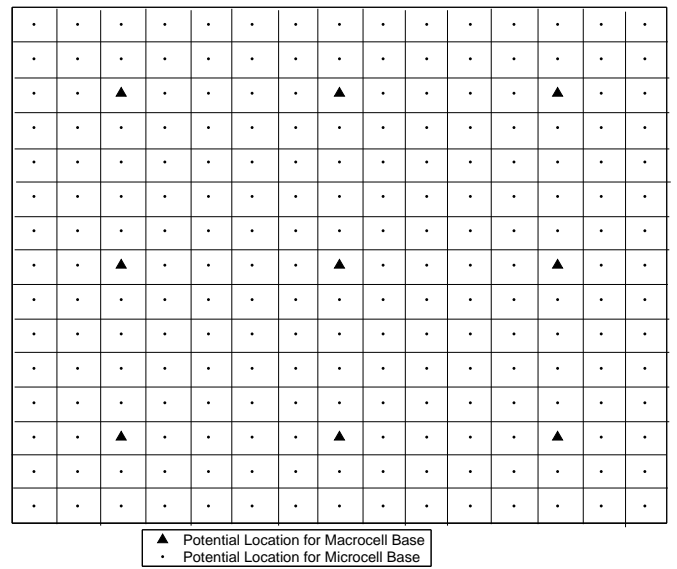,width=3.5in} \caption{An example of the multiple-macrocell/multiple-microcell system, with $m=3$ and $n=5$.} \label{multimulti_fig}
\end{center}
\end{figure}

\begin{figure}[t]
\begin{center}
\epsfig{figure=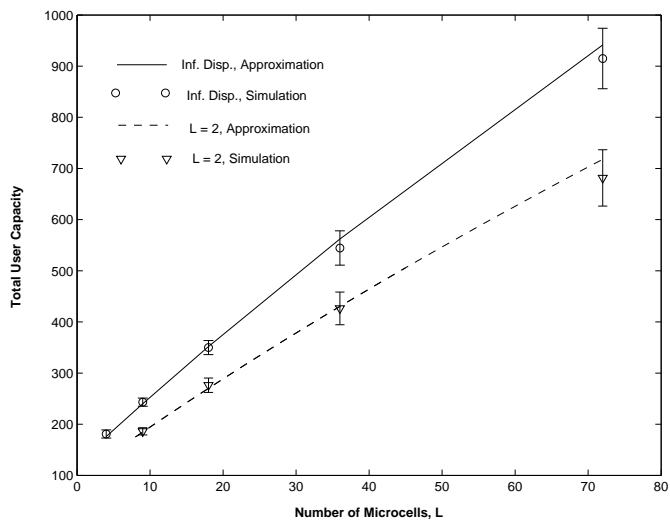,width=3.5in} \caption{Total user capacity as a function of $L$ for the 2-path uniform channel and the infinitely dispersive channel using both simulation and analytical approximation.  The error bars show the standard deviation spread about the mean.  (Unlimited terminal power)} \label{multi_multi_curve}
\end{center}
\end{figure}

We simulated a multiple-macrocell/multiple-microcell system by extending the system studied above.  Specifically (Figure \ref{multimulti_fig}),  there are $m^2$ macrocells and $m^2(n^2-1)$ candidate high density regions (locations for microcell bases).  We randomly select $L$ high density regions from this pool.  We again accrue the average total capacity and its one-standard-deviation spread for the two extreme cases of dispersion, i.e.,  infinite dispersion and $L_p=2$.  The results for $m=3$, $n=5$ are in Figure \ref{multi_multi_curve}.  Also shown are the capacities predicted by (\ref{eq:cap_multi_3}) and (\ref{multi_fading_capacity}) for $M=9$ and $DF=2$.  The approximations closely match simulation points up to at least $L=72$, which translates to roughly eight microcells per macrocell (fill factor of about one third).  Thus, the approximation method is reliable in the domain of hotspot microcells.
\section{Conclusion}
We first examined the effect of transmit power constraints and cell size on the uplink performance of a two-cell two-tier CDMA system.  We then investigated the effect of finite channel dispersion (which causes variable fading of the output signal power) on this capacity.  Analysis was presented to approximate the uplink capacity of {\em any} channel delay profile using the uniform multipath channel, both with and without terminal power constraints.  Finally, we presented an approximation to the uplink user capacity in larger multicell systems with finite dispersion and demonstrated its accuracy using extensive simulation.
\\
\\
While a particular set of system parameters was used in obtaining our numerical results, the analytical approximation methods, and their confirmation via simulations, are quite general.  What we have shown here, first, is that two-tier uplink user capacity can be reliably estimated in a two-cell system for arbitrary pole capacity, channel delay profile, shadow fading variances, transmit power limit, and cell size.  Moreover, under realistic conditions on transmit power and cell size ($F>1$), the method can be extended to the case of multiple macrocells with multiple microcells.  Thus, we have developed simple, general and accurate approximation methods for treating a long-standing problem in two-tier CDMA systems.

\section*{Acknowledgment}
We thank the editor, Dr. Halim Yanikomeroglu, for his helpful
suggestions on expanding the scope of our study.
\bibliographystyle{IEEEbib}

\begin{thebibliography}{1}
\bibitem{Shapira}
J.~Shapira,
\newblock ``Microcell engineering in cdma cellular networks,''
\newblock {\em IEEE Trans. Vehic. Technol.}, vol. 43, no. 4, pp.
  817--825; Nov. 1994.
\bibitem{Wuetal}
J.S.~Wu, et al.,
\newblock ``Performance study for a microcell hot spot embedded in
  cdma macrocell systems,''
\newblock {\em IEEE Trans. on Vehic. Technol.},
  vol.~48, no.~ 1, pp. 47--59; Jan. 1999.
\bibitem{Gaytan}
J.~J. Gayt\'{a}n and D.~Mu\ {n}oz Rodr\'{i}guez,
\newblock ``Analysis of capacity gain and ber performance for cdma systems with
  desensitized embedded microcells,''
\newblock in {\em Proc. of International Conf. on Universal Personal Commun. '98}, 1998, vol.~2, pp. 887--891; Oct. 1998.
\bibitem{skishore1}
S. Kishore, et al.,
\newblock ``Uplink capacity in a cdma macrocell with a hotspot microcell:  exact and approximate analyses,''
\newblock {\em IEEE Trans. on Wireless Comm.}, vol. 2,
no. 2, pp. 364--374; Mar. 2003.
\bibitem{skishore2}
S. Kishore, et al.,
\newblock ``Uplink user capacity of a multi-cell cdma system with
hotspot microcells,''
\newblock in {\em Proc. of Vehic. Techn. Conf. S-02}, vol. 2, pp. 992--996; May 2002.
\bibitem{skishore3}
S. Kishore, et al.,
\newblock ``Uplink user capacity of multi-cell cdma system with
hotspot microcells,''
\newblock submitted to {\em IEEE Trans. on Wireless
Comm.}, Sept. 2003, http://www.eecs.lehigh.edu/\char126 skishore/research/inprogress.htm.
\bibitem{skishore4}
S. Kishore, et al.,
\newblock ``Uplink user capacity in a cdma macrocell with a hotspot
microcell:  Effects of finite power constraints and channel dispersion,'' in {\em Proc. of IEEE Globecom}, vol. 3, pp. 1558--1562, Dec. 2003.
\bibitem{skishore5}
S. Kishore, et al.,
\newblock ``Downlink user capacity in a cdma macrocell with a hotspot microcell,''  in {\em Proc. of IEEE Globecom}, vol. 3, pp. 1573--1577, Dec. 2003.
\bibitem{gilhousen}
 K. S. Gilhousen, et al.,
\newblock ``On the capacity of a cellular cdma system,''
\newblock {\em IEEE Trans. on Vehic. Technol.}, vol. 40, no. 2, pp. 303--312; May 1991.
\bibitem{Erceg}
V.~Erceg, et al.,
\newblock ``An empirically based path loss model for wireless channels in
  suburban environments,''
\newblock {\em IEEE J. on Sel. Areas in Comm.}, vol. 17, no.
  7, pp. 1205--1211; July 1999.
\bibitem{rap}
T. S. Rappaport,
\newblock {\em Wireless Communications:  Principles and Practice},
Prentice Hall, 1996; Chapter 3.
\bibitem{thesis}
S. Kishore,
\newblock {\em Capacity and Coverage in Two-tier Cellular CDMA Networks},
Ph.D. Thesis, Department of Electrical Engineering, Princeton University; Jan. 2003.
\bibitem{3gpp}
``Deployment Aspects,'' 3GPP Specifications, TR 25.943 V4.0.0, (2001-2006), 2001.
\bibitem{lola}
A. Awoniyi, et al., \newblock ``Characterizing the orthogonality factor in wcdma downlink,'' \newblock {\em IEEE Trans. on Wireless Comm.}, vol. 2, no. 4, pp. 611--615; July 2003.
\end{thebibliography}

\begin{table}[h]
\begin{center}
\begin{tabular}{|c|c||c|c|}
\hline $W/R$ & 128 & $h_m$ & 1.5 m
\\ $\Gamma_M$ & 7 dB & $\Gamma_\mu$ & 7 dB
\\ $h_M$ & 60 m & $h_\mu$ & 9 m
\\ $b_M$ & 100 m & $b_\mu$ & 100 m
\\ $H_M$ & $10 H_\mu$ & $D$ & 300 m
\\ $\sigma_M$ & 8 dB & $\sigma_\mu$ & 4 dB
\\ $s$ & 200 m & $S$ & 1 km \\
\hline
\end{tabular}
\\
\caption{System parameters used in simulation.}
\end{center}
\label{sys_param}
\end{table}

\vspace{1in}
\begin{biography}[{\includegraphics[width=1in,height
=1.25in,clip,keepaspectratio]{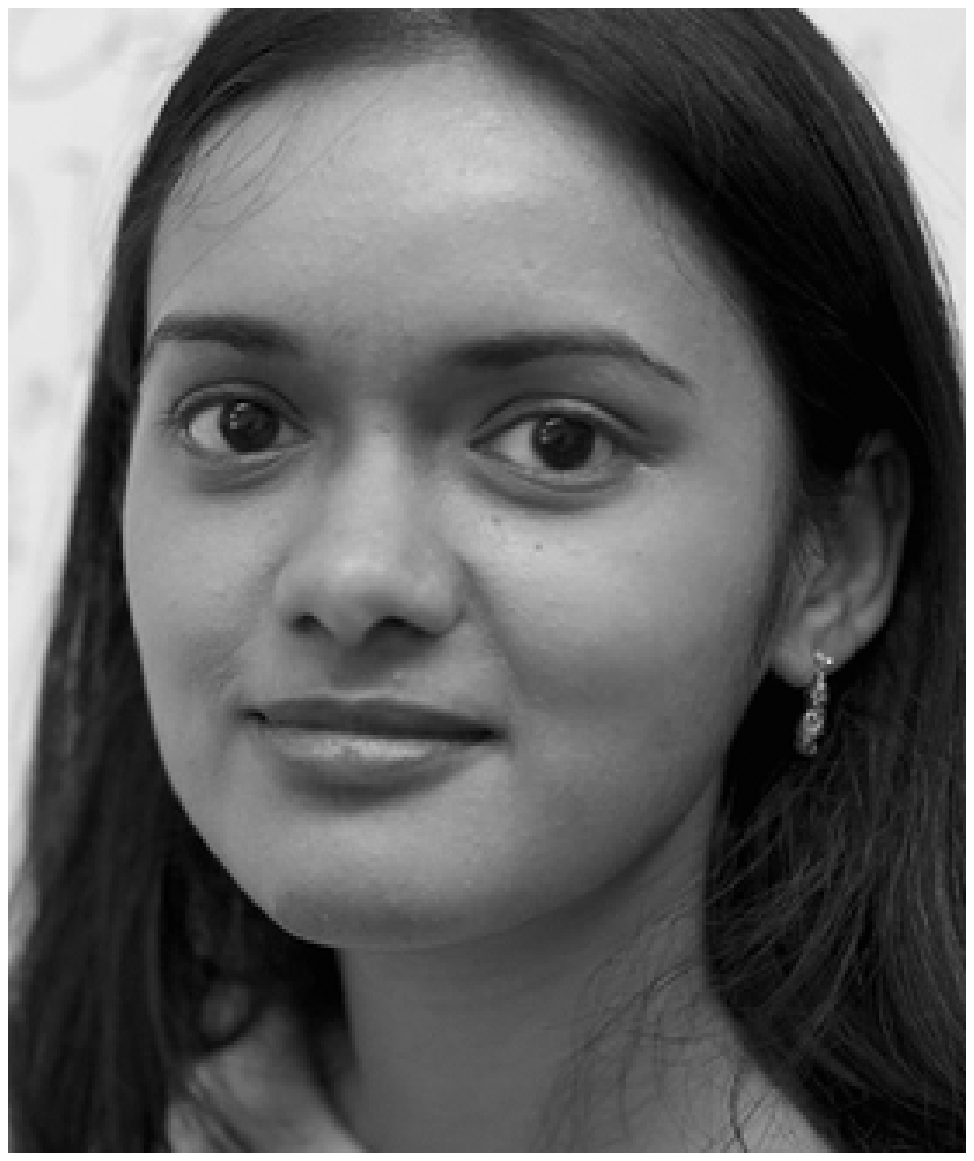}}]{Shalinee Kishore}
      received the B.S. and M.S. degrees in Electrical Engineering
      from Rutgers University in 1996 and 1999, respectively, and
      the M.A. and Ph.D. degrees in Electrical Engineering from
      Princeton University in 2000 and 2003, respectively.

Dr. Kishore is an Assistant Professor in the Department of Electrical and
Computer Engineering at Lehigh University.  From 1994 to 2002,
she has held numerous internships and consulting positions at AT\&T,
Bell Labs, and AT\&T Labs-Research.  She is the recipient of the
National Science Foundation CAREER Award, the P.C. Rossin Assistant
Professorship, and the AT\&T Labs Fellowship Award.  Her research
interests are in the areas of communication theory, networks, and signal
processing, with emphasis on wireless systems.
\end{biography}
\begin{biography}[{\includegraphics[width=1in,height
=1.25in,clip,keepaspectratio]{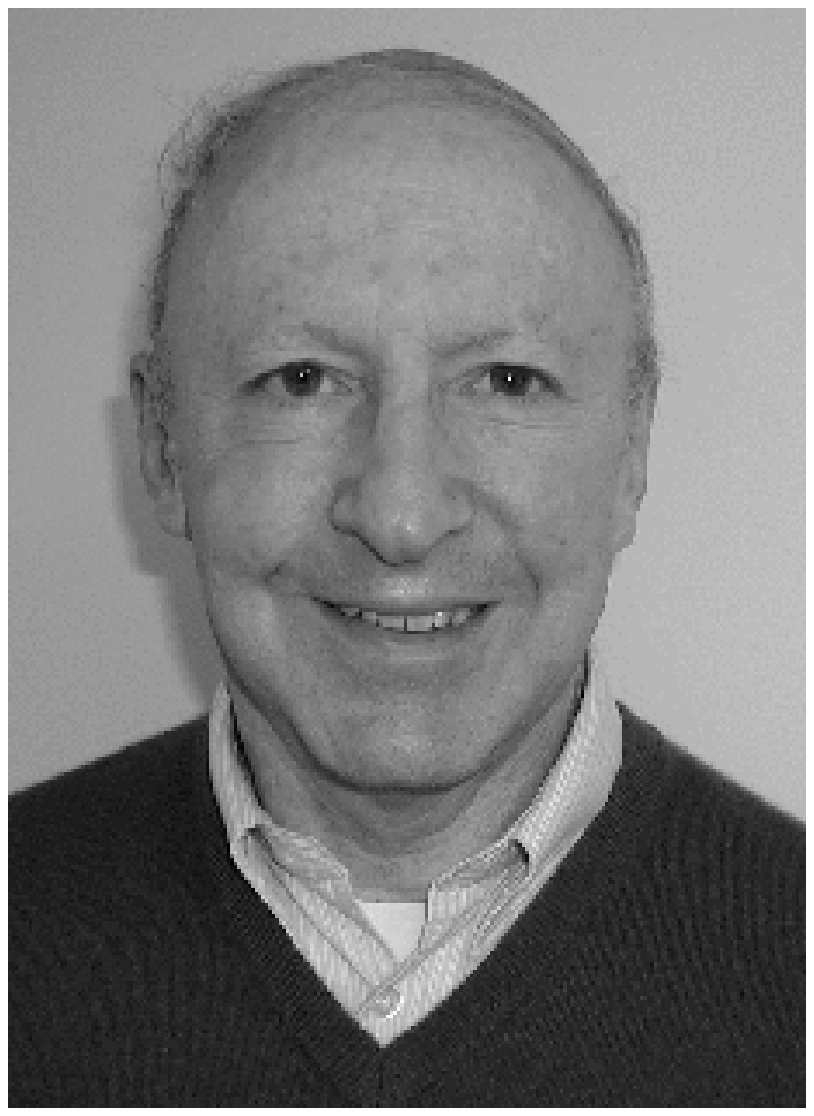}}]{Larry Greenstein}
      (M'59-SM'80-F'87-LF'02) received the B.S., M.S., and
      Ph.D. degrees in Electrical Engineering from Illinois Institute
      of Technology, Chicago, in 1958, 1961, and 1967, respectively.

From 1958 to 1970, he was with IIT-Research Institute, Chicago, IL,
working on radio-frequency interference and anti-clutter airborne
radar.  He joined Bell Laboratories, Holmdel, NJ, in 1970.  Over a
32-year AT\&T career, he conducted research in digital satellites,
point-to-point digital radio, lightwave transmission techniques, and
wireless communications.  For 21 years during that period (1979-2000),
he led a research department renowned for its contributions in these
fields.  His research interests in wireless communications have
included measurement-based channel modeling, microcell system design
and analysis, diversity and equalization techniques, and system
performance analysis and optimization.  Since April 2002 he has been a
research professor at Rutgers WINLAB, Piscataway, NJ, working in the
areas of ultra-wideband systems, sensor networks, relay networks and
channel modeling. Dr. Greenstein is an AT\&T Fellow and a
member-at-large of the IEEE Communications Society Board of
Governors. He has been a guest editor, senior editor and editorial
board member for numerous publications.
\end{biography}

\begin{biography}[{\includegraphics[width=1in,height
=1.25in,clip,keepaspectratio]{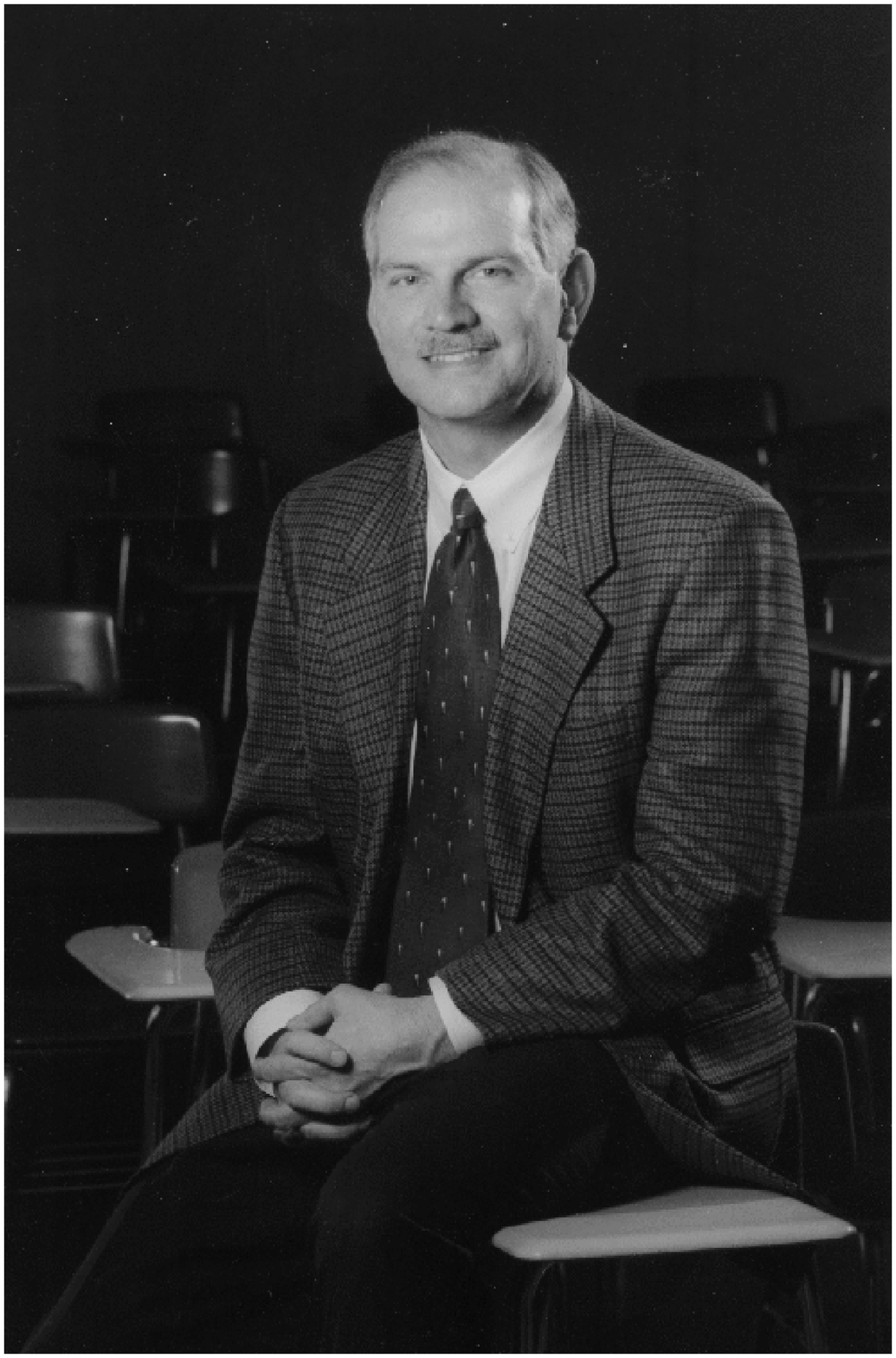}}]{H. Vincent Poor}
(S'72,M'77,SM'82,F'97)
      received the Ph.D. degree in EECS from Princeton University in
      1977.  From 1977 until 1990, he was on the faculty of the
      University of Illinois at Urbana-Champaign.  Since 1990 he has
      been on the faculty at Princeton, where is the George Van Ness
      Lothrop Professor in Engineering.  Dr. Poor's research interests
      are in the area of statistical signal processing and its
      applications in wireless networks and related fields.  Among his
      publications in these areas is the recent book {\em Wireless
      Networks:  Multiuser Detection in Cross-Layer Design}
      (Springer:New York, NY, 2005).

Dr. Poor is a member of the National Academy of Engineering, and is a
      Fellow of the Institute of Mathematical Statistics, the Optical
      Society of America, and other organizations.  In 1990, he served
      as President of the IEEE Information Theory Society, and in
      1991-92 he was a member of the IEEE Board of Directors.  He is
      currently serving as the Editor-in-Chief of the {\em IEEE
      Transactions on Information Theory}.  Recent recognition of his
      work includes the Joint Paper Award of the IEEE Communications
      and Information Theory Societies (2001), the NSF Director's
      Award for Distinguished Teaching Scholars (2002), a Guggenheim
      Fellowship (2002-2003), and the IEEE Education Medal (2005).
\end{biography}

\begin{biography}[{\includegraphics[width=1in,height
=1.25in,clip,keepaspectratio]{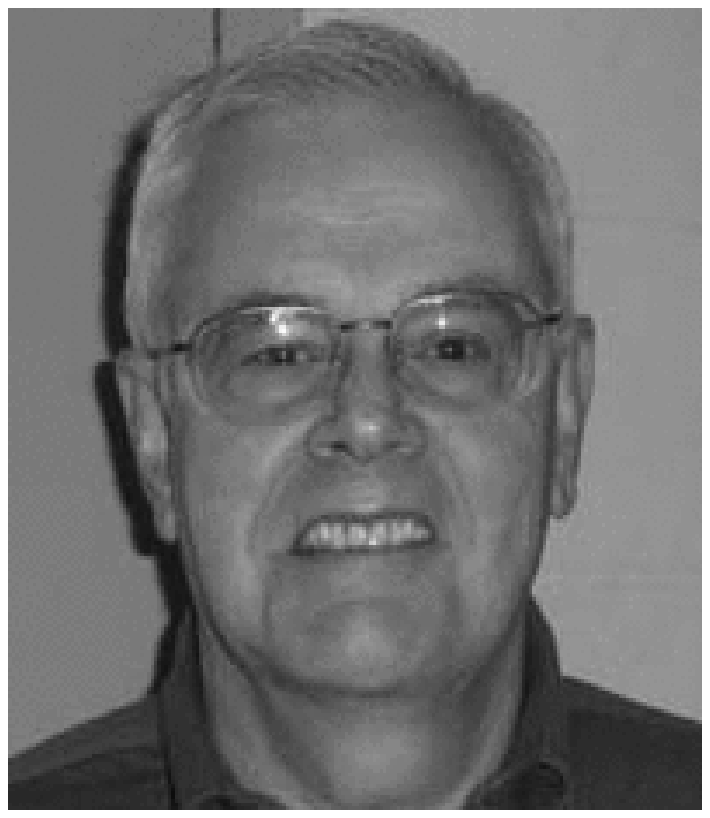}}]{Stuart Schwartz} received the
      B.S. and M.S. degrees from M.I.T. in 1961 and the Ph.D.  from
      the University of Michigan in 1966.  While at M.I.T. he was
      associated with the Naval Supersonic Laboratory and the
      Instrumentation Laboratory (now the Draper Laboratories).
      During the year 1961-62 he was at the Jet Propulsion Laboratory
      in Pasadena, California, working on problems in orbit estimation
      and telemetry.  During the academic year 1980-81, he was a
      member of the technical staff at the Radio Research Laboratory,
      Bell Telephone Laboratories, Crawford Hill, NJ, working in the
      area of mobile telephony.

He is currently a Professor of Electrical Engineering at Princeton
University.  He was chair of the department during the period 1985-1994, and
served as Associate Dean for the School of Engineering during the period July
1977-June 1980.  During the academic year 1972-73, he was a John S.
Guggenheim Fellow and Visiting Associate Professor at the department of
Electrical Engineering, Technion, Haifa, Israel. He has also held visiting
academic appointments at Dartmouth, University of California, Berkeley, and
the Image Sciences Laboratory, ETH, Zurich.  His principal research interests
are in statistical communication theory, signal and image processing
\end{biography}
\end{document}